\documentclass[a4paper]{article}
\usepackage{INTERSPEECH2021}
\usepackage{cite, url}
\usepackage{multicol, multirow}
\usepackage{hyperref}
\usepackage{xcolor}
\usepackage{arydshln}
\usepackage{tabularx}
\newcommand{\newpara}[1]{\vspace{8pt}\noindent\textbf{#1}}
\newcolumntype{Y}{>{\centering\arraybackslash}X}
\hypersetup{
    colorlinks=true,
    linkcolor=purple,
    filecolor=magenta,      
    urlcolor=purple,
}
\title{Three-class Overlapped Speech Detection using a\\
Convolutional Recurrent Neural Network}
\name{Jee-weon Jung, Hee-Soo Heo, Youngki Kwon, Joon Son Chung, Bong-Jin Lee}
\address{
  Naver Corporation, South Korea}
\email{jeeweon.jung@navercorp.com}
\begin{document}
\maketitle
\begin{abstract}
  In this work, we propose an overlapped speech detection system trained as a three-class classifier. 
  Unlike conventional systems that perform binary classification as to whether or not a frame contains overlapped speech, the proposed approach classifies into three classes: non-speech, single speaker speech, and overlapped speech.
  By training a network with the more detailed label definition, the model can learn a better notion on deciding the number of speakers included in a given frame.
  A convolutional recurrent neural network architecture is explored to benefit from both convolutional layer's capability to model local patterns and recurrent layer's ability to model sequential information. 
  The proposed overlapped speech detection model establishes a state-of-the-art performance with a precision of 0.6648 and a recall of 0.3222 on the DIHARD II evaluation set, showing a 20\% increase in recall along with higher precision. 
  In addition, we also introduce a simple approach to utilize the proposed overlapped speech detection model for speaker diarization which ranked third place in the Track 1 of the DIHARD III challenge. 
\end{abstract}

\section{Introduction}
\label{sec:intro}
  % task definition+application
  Overlapped speech detection (OSD) is a task that estimates onsets and offsets of segments (i.e., a small part of an audio clip) within an audio clip (i.e., utterance, session, conversation as a whole) where more than one speaker is speaking simultaneously. 
  OSD can be used for various applications, among which is speaker diarization  where OSD is reported to play a critical role~\cite{Boakye2008OverlappedMeetings, Charlet2013ImpactDebates,Ben-Harush2009FrameDiarization,Boakye2008TwosSpeech,Yousefi2020Frame-BasedNetworks,Andrei2017DetectingLearning,Diez2018BUT2018,Yella2014OverlappingConversations,Wrigley2005SpeechAudio}. 
  Without an OSD system, the performance of a speaker diarization system degrades, proportionally to the existing overlaps in a dataset, because the model conclusively misses the second speaker in the overlapped speech region. 

  % related works
  Before the recent advances in deep neural networks (DNNs), hidden Markov models and Gaussian mixture models were widely used to build OSD systems~\cite{Martin2011TheDiarization, Ben-Harush2009EntropyDiarization.}. 
  Martin \textit{et al.} shows that the hidden Markov model can successfully perform OSD where each state is defined using a Gaussian mixture model. 
  In recent works, OSD is treated as a sequence labeling task where convolutional neural networks (CNNs) or recurrent neural networks (RNNs) are trained as a binary classifier~\cite{Cornell2020DetectingScenarios,Yousefi2020Frame-BasedNetworks, Geiger2013DetectingNetworks, Bullock2020Overlap-AwareDetection}. 
  Bullock \textit{et al.}~\cite{Bullock2020Overlap-AwareDetection} adopts two bidirectional long short-term memory recurrent layers~\cite{Hochreiter1997LongMemory} and demonstrates state-of-the-art performance on the DIHARD II evaluation set. 

  % proposal-OSD
  In this study, we make two proposals on top of existing OSD researches. 
  First, we train the OSD model as a three-class classifier instead of a binary classifier. 
  Before the widespread use of DNNs in OSD, hidden Markov models were used for OSD\cite{Wrigley2005SpeechAudio} where OSD was modeled via transitions between three independent states: non-speech, single speaker speech, and overlapped speech. 
  However, when using DNNs for OSD, only two classes (overlapped and non-overlapped) are defined. 
  We argue that the DNN can also benefit from defining more specific classes based on additional motivations addressed in Section \ref{ssec:3class}. 
  Second, we use a convolutional recurrent neural network (CRNN) architecture to merit from both architecture's features. 
  Although studies adopting CRNN show state-of-the-art performance in other sequence labeling tasks such as sound event detection~\cite{Pi2019PolyphonicStrategy}, existing OSD models adopt variants of either a CNN or a RNN. 
  We assume that OSD models can be also improved by adopting a CRNN architecture, benefiting from both CNN and RNN's unique modeling capability. 
  In the upcoming sections, we explain and experimentally show that both proposals increase the performance of an OSD model and establish state-of-the-art performance. 

  % proposal-OSD-SD algorithm
  Exploiting the developed OSD model, we also introduce a simple algorithm for detecting simultaneously speaking speakers in a diarization task. 
  It adds two sub-processes using an OSD model to the speaker diarization pipeline. 
  First, we add a sub-process that further segments speech activity detection (SAD) module's output where the onsets and offsets of overlapped segments are adjusted using the SAD result. 
  Second, we provide an additional label to the overlapped segment using the second ranked speaker based on the clustering result of speaker diarization. 
  This method works independently from the mainstream speaker diarization pipeline, making it simple to utilize yet effective, ranking third in the DIHARD III challenge Track 1 core evaluation~\cite{Ryant2020ThirdPlan}. 

\begin{figure}[t!]
  \centering
  \includegraphics[width=\linewidth]{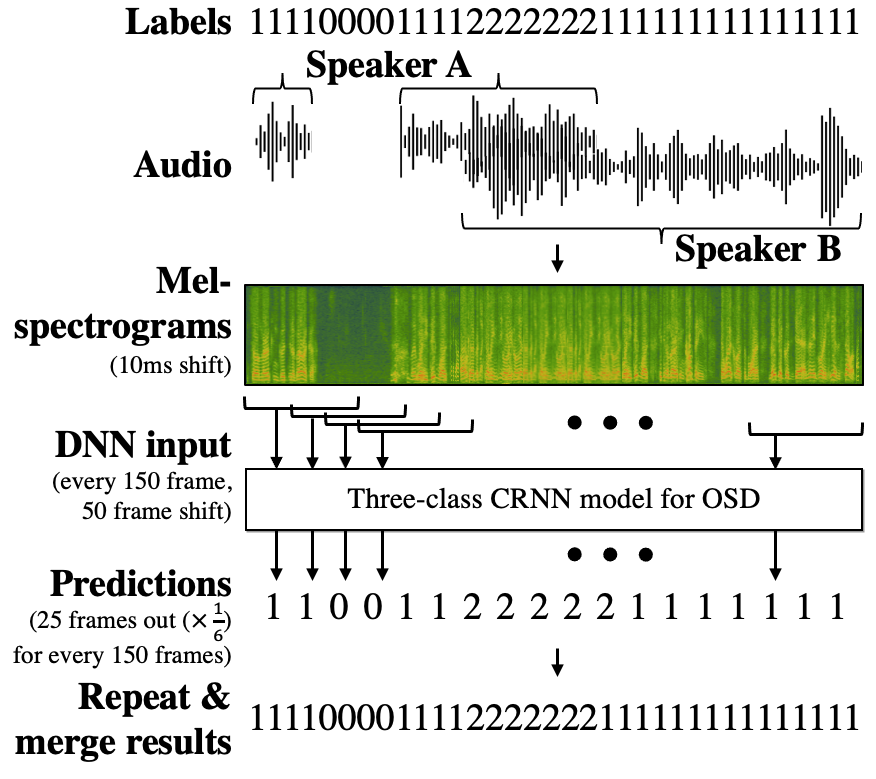}
  \caption{
    Overall pipeline of the proposed CRNN-based three-class OSD system in the evaluation phase. 
    An input audio clip is frame-wisely classified into one of the three classes where 0,1 and 2 represent non-speech, single speaker speech, and overlapped speech, respectively.
    To process the entire audio clip longer than the defined sequence length of 1.5 seconds, every 150 frames are input to the model with 50 frames shift and model outputs are combined.
  }
  \label{fig:osdframework}
\end{figure}

\section{Overlapped speech detection}
\label{sec:OSD}
  % notation
  Defined as a sequence labeling task, an OSD model inputs $\mathbf{x}=[\mathbf{x_1}, \mathbf{x_2}, ..., \mathbf{x_t}], \mathbf{x_t} \in \mathbb{R}^{f}$ where $t$ is the length of input sequence and $f$ refers to the dimensionality of each element. 
  Acoustic features (e.g., mel-spectrograms or mel frequency cepstral coefficients) are widely chosen as the input feature. 
  Corresponding label $\mathbf{y}=[y_1, y_2, ..., y_t], y_t \in \mathbb{R}$ is used to guide the training where 0 refers to \textit{not overlapped} and 1 refers to \textit{overlapped} (can be the other way).
  The output layer of the model is either one node, trained using a binary cross-entropy loss function, or two nodes, trained using a categorical cross-entropy loss function. 

  % framework
  The overall framework of our proposed OSD model when conducting inference using an input audio clip is illustrated in Figure \ref{fig:osdframework}. 
  Note that for conventional DNN-based OSD models, labels are either 0 or 1 as mentioned earlier, but for the proposed three-class OSD model, labels are either 0, 1, or 2 indicating non-speech, single speaker speech, and overlapped speech.
  We adopt mel-spectrograms as the input feature, using a 25ms window and 10ms shift size. 
  The OSD model inputs 150 frames ($\approx$ 1.5s) at a time with 50 frames shift. 
  The node of the output layer that indicates overlapped speech is used. 
  Except for the first and the last 100 frames, the sum of model output is divided by three accounting for the multiple predictions caused by shifting the input. 
  To compensate for the decrease in the length of sequence affected by pooling layers, the values are duplicated.
  The main contributions of this paper, three-class classification and the CRNN architecture, are described in Sections \ref{ssec:3class} and \ref{ssec:CRNN} respectively.
  
\begin{table}[t]
 \caption{Architecture of the proposed CRNN for three-class OSD. A mel-spectrogram of size (150, 128, 1) is input to the DNN where 150 refers to the length of sequence (1.5s), 128 is the number of mel bins, and 1 is the number of channels. Two numbers in the Conv bracket refers to filter length and stride size (BN: batch normalization, SE: squeeze-excitation). Average is applied to the mel bin dimension, before the GRU layer.}
  \centering
  \label{tab:DNN_arch}
  \begin{tabular}{l c c}
  \toprule
  \textbf{Layer} & \textbf{Input:(\#seq,\#mel,\#filt)} & \textbf{Output shape}\\
  \toprule
  CNN & 
    $\left \{
      \begin{tabular}{c}
      Conv(3,1) \\
      BN\\
      ReLU\\
      Conv(3,1)\\
      BN\\
      ReLU\\
      SE\\
      AveragePool\\
      \end{tabular}
    \right \}$
    $\times$3
    
  & (25, 32, 128)\\
  \midrule
  Average & Average & (25,128)\\
  \midrule
  GRU & bi-GRU(256)$\times$2 & (512,)\\
  \midrule
  Linear & 
    $\left \{
      \begin{tabular}{c}
      FC(256) \\
      Dropout(0.5)\\
      LeakyReLU\\
      Linear(3)\\
      \end{tabular}
    \right \}$
  & (3,)\\
  \bottomrule
  \end{tabular}
\end{table}

\subsection{OSD as a three-class classification}
\label{ssec:3class}
  Conventional OSD systems using DNNs are trained to perform binary classification because the objective at the inference phase is to decide whether overlapped speech exists for each frame. 
  It is a widely accepted straightforward approach, training the model to perform as one wants. 
  However, by training the OSD model as a two-class classifier, both non-speech and single speaker segments are labeled as \textit{not overlapped}. 
  We assume that grouping non-speech and single speaker speech into not overlapped, regardless of the existence of human voice, may not be ideal in two aspects. 
  First, two small clusters may be created within not overlapped class, diminishing the OSD model's generalization.
  Second, training OSD as a two-class classification task intensifies class imbalance. 
  For instance, suppose there exists a sample in the train set that has 40 non-speech frames, 100 single speaker speech frames, and 10 overlapped speech frames.
  Training it with conventional two-class results in 14:1 imbalance ratio, but with three-class training, it becomes 4:10:1. 
  Although manual weight can be set for each class, it is difficult to account for the single speaker segments.
  
  Thus, we propose to train the OSD model as a three-class classifier where the classes are non-speech, single speaker speech, and overlapped speech.
  This is inspired by studies~\cite{Martin2011TheDiarization, Ben-Harush2009EntropyDiarization.} that use hidden Markov models where single speaker speech and non-speech are separately defined as states. 
  By dividing not overlapped class into single speaker speech and non-speech classes, we argue that not only the problem occurred by grouping different categories may decrease, but also the DNN can leverage from additional information. 
  It has commonalities in terms of motivation with the multitask learning~\cite{Caruana1997MultitaskLearning} framework where more specific and concrete information is additionally provided to help learning. 
  Also, precise class balancing using loss weights becomes possible.
  The proposed three-class scheme has no additional requirements when performing inference. 
  One can simply use the node of the output layer that indicates overlapped speech (before or after applying softmax non-linearity) as the model's score for the given input frame. 
%   In addition, by revising the model into a three-class classifier, the model can be also used for SAD by using the output layer's node denoting non-speech. 

\subsection{CRNN architecture}
\label{ssec:CRNN}
  % 1. CRNN motivation
  Most preceding studies that use DNN-based models for the OSD task either utilize a CNN or a RNN~\cite{Cornell2020DetectingScenarios,Yousefi2020Frame-BasedNetworks, Geiger2013DetectingNetworks, Bullock2020Overlap-AwareDetection}. 
  However, we propose to use a CRNN architecture with three convolution blocks and two gated recurrent unit (GRU) layers. 
  Three motivations led us to propose and use the CRNN architecture. 
  First, CNNs and RNNs have different aspects that the OSD model can benefit from. 
  CNNs are known for their capability of modeling local patterns and RNNs are widely used for their capability of modeling sequential data. 
  We argue that by using a CRNN architecture, advantages of both CNNs and RNNs can be exploited. 
  Second, decreasing the length of sequence via pooling layers included in CNN will facilitate the training of GRU layers. 
  In addition, decreasing the sequence length will give a smoothing effect on the model's output. 
  For example, according to our configuration when 150 frames are input, the CRNN model will output 25 frames where each output frame represents six input frames. 
  Third, a CRNN architecture shows state-of-the-art performance in sound event detection~\cite{Pi2019PolyphonicStrategy} which is also a sequence labeling task that predicts onsets and offsets of various predefined sound events. 
  
  % 2. architecture overall
  Table \ref{tab:DNN_arch} denotes the proposed CRNN architecture used for OSD. 
  Three convolution blocks are used where the length of a sequence is divided by 6, followed by an average on the mel bin dimension. 
  A squeeze-excitation~\cite{Hu2018Squeeze-and-ExcitationNetworks} is applied to each block. 
  Details on how the length of the sequence and the number of mel bins are decreased are addressed in Section \ref{ssec:exp_config}. 
  The output of CNN is input to two bidirectional GRU layers that output sequences. 
  Each frame is then classified among the three classes and lastly duplicated to match the length of the input sequence. 
  
\begin{figure}[t!]
  \centering
  \includegraphics[width=0.8\linewidth]{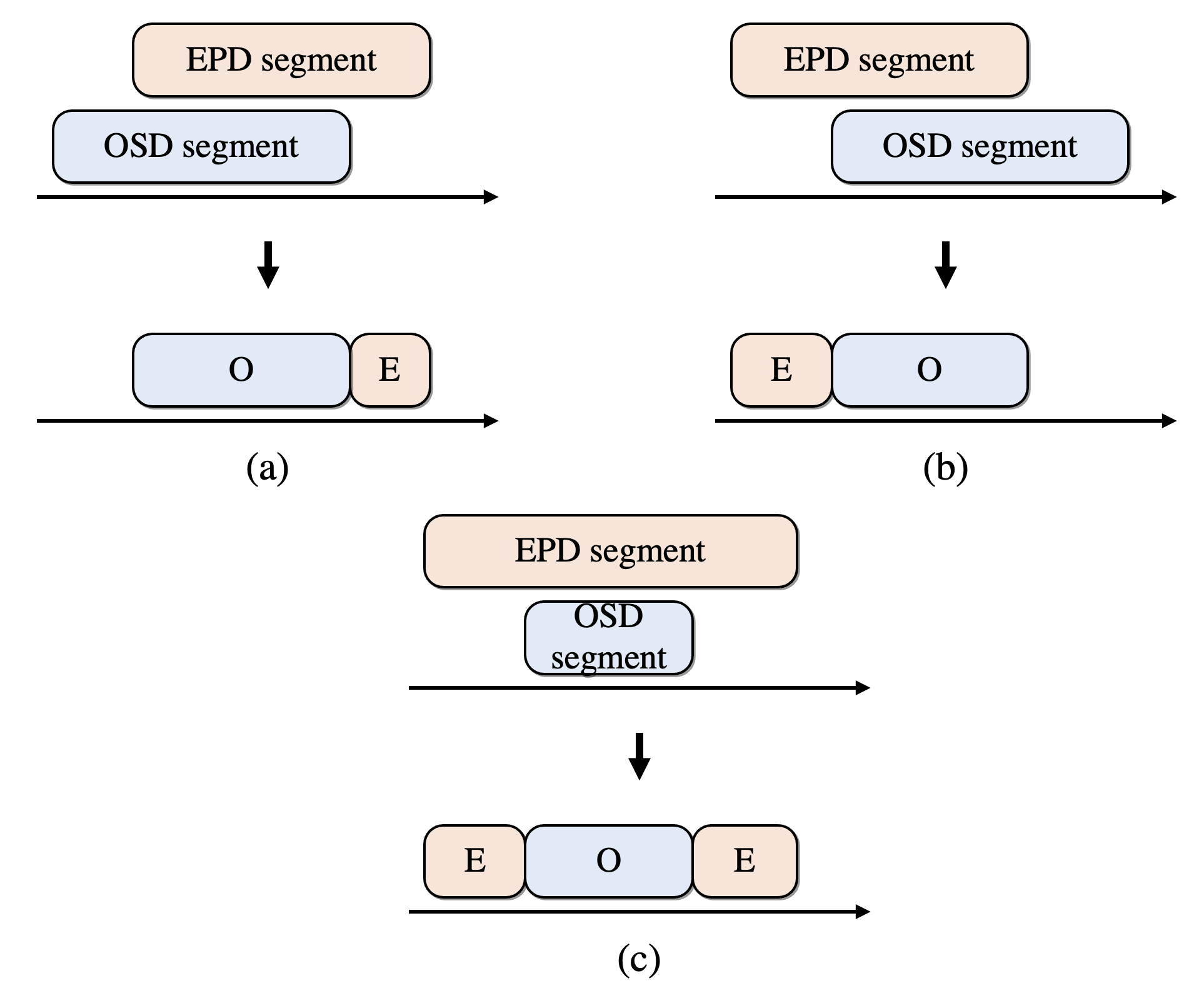}
  \caption{Further segmentation of SAD segment using OSD result.}
  \label{fig:osd_segmenting}
\end{figure} 

\subsection{Overlap and sampling rate augmentation}
\label{ssec:overlapAug}
  We adopt two data augmentation techniques on top of widely used noise and reverberation augmentation~\cite{Povey2011TheToolkit, Heo2020Clova2020}. 
  First, we generate augmented overlap speech samples using two single speaker segments alike the method introduced in Bullock \textit{et al.}~\cite{Bullock2020Overlap-AwareDetection}. 
  It diminishes the class imbalance problem and generates sufficient overlapped speech samples; most datasets that can be used for training an OSD model have less than 10\% overlapped speech regions. 
  Second, we augment training samples by downsampling an audio clip and then upsampling back. 
  We use this augmentation to cope with telephone conversation audio clips that have been upsampled in the test phase. 
  Through experiments, we empirically find that this augmentation mitigates performance degradation on 8kHz sampling rate audio clips that exist in the evaluation dataset.

\section{Overlapped segment clustering in speaker diarization}
\label{sec:OSD_SD}
  % SD using OSD: why OSD is critical for SOTA SD
  Speaker diarization is a widely used application of an OSD model. 
  Processing overlapped segments have been recognized as a key for developing state-of-the-art speaker diarization systems. 
  This is because, in the conventional diarization systems that cannot assign multiple labels to one segment, 
  the ratio of overlapped speech directly leads to an increase in missed speech, increasing diarization error rate (DER). 
  
  % segmenting SAD segments using OSD
  In order to utilize the proposed OSD model for speaker diarization, we introduce two simple steps that can be applied on top of the existing speaker diarization pipeline. 
  First, we use the OSD model to further divide SAD results into single speaker segment and overlapped speech segment. 
  Regions where both the OSD model and the SAD model detects are assigned as overlapped speech segment.
  Figure \ref{fig:osd_segmenting} illustrates three possible cases of the OSD model further segmenting the SAD segment. 
  If the OSD onset is earlier than the SAD onset, SAD onset is used as the onset of an overlapped segment and OSD offset is used as modified SAD segment onset (Figure \ref{fig:osd_segmenting}-(a)). 
  If the OSD offset positions later than the SAD offset, SAD offset is used as an overlapped segment's offset and SAD offset is replaced by OSD onset (Figure \ref{fig:osd_segmenting}-(b)). 
  When the OSD segment exists in the middle of an SAD segment, it is divided into three segments (Figure \ref{fig:osd_segmenting}-(c)). 
  
  % adding the second speaker using the existing clustering approach
  The other application of the OSD model for speaker diarization occurs after the clustering phase, in a similar manner with \cite{Bullock2020Overlap-AwareDetection}. 
  When performing clustering, there exists a speaker posterior matrix where speakers are aligned from the most likely to least likely for each segment. 
  For overlapped segments, we simply allocate the second most likely speaker as an additional label. 
  This approach is simple, easy to implement, and can be applied to the result of various clustering methods (e.g., spectral and agglomerative hierarchical clustering~\cite{Luxburg2007AClustering,Ning2006ADiarization,Han2008StrategiesDiarization}) that can derive the second most likely speaker for each segment. 
  However, note that the limitation of this approach is not being able to detect when three or more speakers are uttering simultaneously. 
  
  % Clova SD model
  Aforementioned approach is applied to the Clova team's speaker diarization model~\cite{Heo2021NAVERCHALLENGE}. 
  This model uses a reference SAD following the rules of Track 1 of the DIHARD III challenge. 
  For speaker embedding extraction, a variant of ResNet34V2 in voxceleb\_trainer\footnote{\url{https://github.com/clovaai/voxceleb\_trainer}} is used. 
  Speaker embeddings are then fed into two feature enhancement modules, followed by spectral clustering and agglomerative hierarchical clustering~\cite{Luxburg2007AClustering,Ning2006ADiarization}. 
  Further information regarding this speaker diarization system including feature enhancement techniques can be found in~\cite{Heo2020Clova2020}.

\section{Experiments}
\label{sec:exp}

\subsection{Datasets}
\label{ssec:db}
We use four datasets to train the proposed OSD model: AMI Meeting corpus \cite{Carletta2007UnleashingCorpus}, DIHARD I development set\cite{Ryant2018FirstPlan}, DIHARD II development set \cite{Ryant2019TheBaselines}, and VoxConverse dataset \cite{Chung2020SpotWild}. 
The AMI corpus includes 100 hours of meeting conversations. 
DIHARD I and II development sets include 19 and 23 hours of data from diverse domains respectively. 
The VoxConverse dataset consists of approximately 20 hours of conversation from YouTube videos. 
For all datasets, multi-channel audio clips were modified to single-channel and the sample rate of all audio clips was adjusted to 16kHz. 
The DIHARD II evaluation set and the DIHARD III development set~\cite{Ryant2020TheChallenge} are used to report the performance of the proposed OSD model. 
OSD model's performance is reported using precision and recall. 
Speaker diarization result is reported in DER, the summation of false alarm, miss, and confusion. 

\subsection{Configurations and details}
\label{ssec:exp_config}
  We use a 128-dimensional log mel-spectrograms (i.e., filterbank energies) as input to the DNN with a window size of 25ms and 10ms shift. 
  Pre-emphasis is applied before applying Fourier transforms and mean normalization is applied to the extracted log mel-spectrograms. 
  
  \newpara{Data augmentation.}
  A copy of all audio clips with a sampling rate of 16kHz is augmented by downsampling to 8kHz and upsampling back. 
  We also apply data augmentation using reverberation and noise from RIRs and Musan datasets~\cite{Ko2017ARecognition,Snyder2015MUSAN:Corpus}, following existing recipes~\cite{Povey2011TheToolkit,Heo2020Clova2020}. 
  Overlapped speech sample augmentation following the recipe introduced in Bullock \textit{et al.}~\cite{Bullock2020Overlap-AwareDetection} is also applied. 
  
  \newpara{DNN training and inference.}
  We modify the input sample to 150 frames (1.5s) for batch construction when training the CRNN model. 
  An average pooling with size (2,1), (3,2), (1,2) is applied after each convolution block in Table \ref{tab:DNN_arch}, decreasing the length of input sequence and number of frequency bins to 25 frames and 32, respectively. 
  Each output frame is six times duplicated to match the original sequence length. 
  The model is trained 60 epochs using a batch size of 1024, using four Nvidia Tesla P40 GPUs. 
  Adam optimizer with cosine annealing learning rate scheduler is used. 
  To account for class imbalance, we give weights to categorical cross-entropy loss proportionally. 
  At the inference phase, we input 150 frames to the model with 50 frames shift to handle audio clips with varying duration.

\begin{table}[t!]
  \centering
  \caption{Comparison with the reported state-of-the-art OSD model on DIHARD II evaluation set. Performance of the identical model is additionally reported on DIHARD III development set as a reference (DH: DIHARD, \textbf{higher is better for both precision and recall}).}
  \begin{tabularx}{\linewidth}{lYYY}
  \toprule
  System         & Dataset & Precision & Recall\\
%   \midrule
%   \textit{DIHARD2 evaluation set}\\
  \midrule
  Bullock \textit{et al.}\cite{Bullock2020Overlap-AwareDetection} & DH2 eval & 0.6450          & 0.2670\\ 
  \textbf{Ours} & DH2 eval & \textbf{0.6648} & \textbf{0.3222}\\ 
  \midrule
%   \textit{DIHARD3 development set}\\
  \midrule
   \textbf{Ours} & DH3 dev & \textbf{0.9000} & \textbf{0.4609}\\ 
  \bottomrule
  \end{tabularx}
  \label{tab:SOTA}
\end{table}
% \begin{table}[t!]
%   \centering
%   \begin{tabularx}{\linewidth}{lYY}
%   \toprule
%   System         & Precision & Recall\\
%   \midrule
%   \textit{DIHARD2 evaluation set}\\
%   \midrule\midrule
%   Bullock \textit{et al.}\cite{Bullock2020Overlap-AwareDetection} & 0.6450          & 0.2670\\ 
%   \textbf{Ours}                            & \textbf{0.6648} & \textbf{0.3222}\\ 
%   \midrule
%   \textit{DIHARD3 development set}\\
%   \midrule\midrule
%   Bullock \textit{et al.}\cite{Bullock2020Overlap-AwareDetection} & -          & -\\ 
%   \textbf{Ours}                            & \textbf{0.9229} & \textbf{0.3947}\\ 
%   \textbf{Ours}                            & \textbf{0.9000} & \textbf{0.4609}\\ 
%   \bottomrule
%   \end{tabularx}
%   \caption{Comparison with the reported state-of-the-art OSD model on DIHARD2 evaluation set. Performance of the identical model is additionally reported on DIHARD3 development set as a reference.}
%   \label{tab:SOTA}
% \end{table}
% performance in 
% https://oss.navercorp.com/NSASR3/dev5/issues/296#issuecomment-5601448
\begin{table}[t!]
  \centering
  \caption{Comparison experiments with ablations showing the effectiveness of the proposed three-class OSD approach and CRNN architecture. Threshold was set to a point where the system shows precision of 0.9 (\textbf{higher is better for both precision and recall}).}
  \begin{tabularx}{\linewidth}{lYYYYY}
  \toprule
  System  & Precision & Recall\\
  \midrule
%   \textbf{Best} (all combined) & 0.9229 & 0.3947\\ 
  \textbf{Best} (all combined) & 0.9000 & 0.4609\\ 
  \midrule
  w/o three-class training  & 0.9000 & 0.4113\\ 
  w/o CNN & 0.9000 & 0.4318\\
  w/o GRU & 0.9000 & 0.3603\\
  \midrule
  3 GRU layers         & 0.9000 & 0.4365\\
  1 GRU layers         & 0.9000 & 0.3314\\
  \midrule
  w/o sequence compression & 0.9000 & 0.4471\\
  $\frac{1}{3}$ sequence compression & 0.9000 & 0.4397\\
  $\frac{1}{10}$ sequence compression & 0.9000 & 0.4135\\
  \bottomrule
  \end{tabularx}
  \label{tab:3class_CRNN}
\end{table}

\subsection{Results}
\label{ssec:result}
  Table \ref{tab:SOTA} describes the comparison between the existing model and the proposed OSD model applying both three-class training and CRNN architecture. 
  The OSD model proposed in Bullock \textit{et al.} \cite{Bullock2020Overlap-AwareDetection}, which demonstrated state-of-the-art performance in AMI and DIHARD II evaluation set is used as the baseline. 
  Experimental results show that our proposed OSD model establishes a new state-of-the-art performance on the DIHARD II evaluation set. 
  As the DIHARD III challenge has been held recently, we further report the proposed model's performance on its development set. 
  Note that this model is similar to, but performs slightly better than the OSD model we used when participating the DIHARD III challenge~\cite{Heo2021NAVERCHALLENGE}. 
  Hereafter, we report all performances using the DIHARD III development set. 
  
  Table \ref{tab:3class_CRNN} addresses comparison experiments with ablation studies to show the effectiveness of training the OSD model as a three-class classifier and using a CRNN architecture. 
  Using the best performing system in the first row as a pivot, we first remove the individual proposed approach. 
  Then, ablations experiments on the number of GRU layers and the extent of sequence length compression are dealt. 
  All performances are reported using a threshold that shows a precision of 0.9.
  
  Results from the second to the fourth rows show that both three-class training and CRNN architecture successfully improves the OSD model's performance. 
  Thee-class training relatively improved recall by 12\%.
  CRNN architecture was also proven effective where removing GRU layers degraded the performance more.
  Through the fifth and sixth rows, we conclude that the model benefits from multiple GRU layers where using one GRU layer decreased performance the most. 
  The last three rows show that compressing the sequence length using pooling layers, before GRU, is also effective for OSD. 
  Reducing training sequence length was effective than not reducing at all, but significantly decreasing also decreased the performance.

  Table \ref{tab:SD_OSD} reports speaker diarization performance using the proposed OSD model on the DIHARD III Track 1 development set. 
  Performance is reported using DER, equal to the summation of false alarm, miss, and confusion. 
  The speaker diarization system refers to that of the Naver Clova team's submission for the DIHARD III challenge~\cite{Heo2021NAVERCHALLENGE}. 
  OSD model threshold was set to a value that shows a precision of 0.95. 
  
  Experimental results show that further segmentation and allocating additional speaker labels using the developed OSD model successfully lowers the DER of speaker diarization, decreasing the DER from 14.29\% to 13.74\% on the DIHARD III development set. 
  Miss, which the OSD model should reduce, decreased to 9.40\% showing relative improvement of 8.73\%, at the cost of an increase in false alarm by 0.09 which was caused by the OSD model predicting single speaker regions as overlapped. 
  Our system shows 32\% lower DER than the challenge baseline and are comparable to the challenge winning system. 
  
\begin{table}[t!]
  \centering
  \caption{Effectiveness of the OSD model when applied to the speaker diarization pipeline using the DIHARD III development set track 1 configuration (FA: false alarm, Conf: confusion, \textbf{lower is better for all four metrics}).}
  \begin{tabularx}{\linewidth}{lYYYYY}
  \toprule
  OSD configuration  & DER & FA & Miss & Conf\\
  \midrule
  w/o OSD   & 14.29 & 0.43 & 10.30 & 3.56\\
  w/ OSD    & 13.74 & 0.52 & 9.40 & 3.81\\%p0.9, r0.4609
  \midrule
  Challenge Baseline~\cite{Ryant2021TheChallenge} & 20.25 & - & - & -\\
  Challenge Winner~\cite{Wang2021USTC-NELSLIPChallenge} & 13.30 & - & - & -\\
%   w/ OSD    & 13.93 & 0.62 & 9.00 & 4.31\\%p0.9, r0.4609
%   w/ OSD    & 14.03 & 0.82 & 8.70 & 4.51\\%p0.8, r0.4609
  
%  None               & a & b & c & d \\ 
%  Segment            & 2.24     & 2.20    & 2.14    \\ 
%  Add label           & 2.09    & 1.62    & 1.17    \\ 
%  Segment\&Add label  & \textbf{2.06}     & \textbf{1.52}    & \textbf{1.09}    \\
  \bottomrule
  \end{tabularx}
%   \caption{Speaker diarization performance using OSD model. `Segment' refers to applying further segmentation on the EPD segment using OSD and `Add label' refers to allocating second best speaker for OSD segments (FA: false alarm, Conf: confusion).}
  \label{tab:SD_OSD}
%   \vspace{-4pt}
\end{table}

% \section{Conclusion and future works}
\section{Conclusion}
\label{sec:conclu}
  We proposed an OSD model that is trained as a three-class classifier dividing the non-overlapped class into non-speech and single speaker speech classes. 
  A CRNN architecture was also proposed to incorporate both CNN and RNN's capability in detecting overlapped speech. 
  Combining two proposals, our OSD model demonstrates state-of-the-art performance on the DIHARD II evaluation set. 
  An approach of using a developed OSD model for speaker diarization was also addressed, where the system that followed this approach ranked third place in DIHARD III Track 1 challenge. 
%   As a future work, we intend to study utilizing the proposed three-class OSD system for SAD. 

\bibliographystyle{IEEEtran}
\clearpage
\bibliography{references_mendeley_jung}
\end{document}